\documentclass{llncs}

\usepackage{cite}
\ifx\pdfoutput\undefined
	\usepackage{graphicx}
\else
	\usepackage[pdftex]{graphicx}
\fi
\graphicspath{{figures/}}
\usepackage{amsmath}
\usepackage{listings}
\usepackage{amsfonts}

\usepackage{xcolor}

\usepackage{amsfonts}
\usepackage{url}

\begin{document}


\title{Role-Oriented Code Generation in an Engine for Solving Hyperbolic PDE Systems}
\titlerunning{Role-Oriented Code Generation in ExaHyPE}

\author{Jean-Matthieu Gallard
\and Lukas Krenz
\and Leonhard Rannabauer
\and Anne Reinarz
\and Michael Bader}

\institute{Department of Informatics, Technical University of Munich
\email{\{gallard,lukas.krenz,leonhard.rannabauer,reinarz,bader\}@in.tum.de} 
}


\maketitle


\begin{abstract}

The development of a high performance PDE solver requires the combined expertise of interdisciplinary teams with respect to 
application domain, numerical scheme and low-level optimization.
In this paper, we present how the ExaHyPE engine facilitates the collaboration of such teams by isolating three roles: 
application, algorithms, and optimization expert. 
We thus support team members in letting them focus on their own area of expertise while integrating their contributions into an HPC production code.

Inspired by web application development practices, ExaHyPE relies on two custom code generation modules, the Toolkit and the Kernel Generator, which follow 
a Model-View-Controller architectural pattern on top of the Jinja2 template engine library.
Using Jinja2's templates to abstract the critical components of the engine and generated glue code, we isolate the application development from the engine.
The template language also allows us to define and use custom template macros that isolate low-level optimizations from the numerical scheme described in the templates.

We present three use cases, each focusing on one of our user roles, showcasing how the design of the code generation modules allows to 
easily expand the solver schemes to support novel demands from applications, 
to add optimized algorithmic schemes (with reduced memory footprint, e.g.), 
or provide improved low-level SIMD vectorization support.

\end{abstract}



\section{Introduction}
ExaHyPE (``An Exascale Hyperbolic PDE Engine'', \url{www.exahype.eu}) is an EU Horizon 2020 project to develop 
an exascale-ready general solver for hyperbolic systems of partial differential equations (PDEs). 
Intended as an \emph{engine} (as in ``game engine''), it concentrates on a dedicated numerical scheme and 
on a fixed mesh infrastructure, but provides flexibility in the PDE system to be solved~\cite{Reinarz:2019}.
Its mission statement is ``to enable medium-sized interdisciplinary research teams to realize
extreme-scale simulations of grand challenges modelled by hyperbolic conservation laws''.
We anticipate (and have observed in our project) that such endeavours progress in phases:
from first attempts to implement the desired PDE model in the engine (realizing simple analytic setups) 
via application-oriented benchmark setups (to validate numerical schemes) towards 
large-scale demonstrator scenarios that establish the viability of the engine to tackle grand challenges. 
We also need to envisage that successful demonstrators shall be further developed into production codes 
or even services.

Orthogonal to the requirements of designing more and more complex applications, we are facing the challenges of 
upcoming exascale architectures. The engine needs to take into account architecture-specific optimizations, which, 
however, again need to be tailored to the specific PDE system and variants of the numerical schemes. 
A common approach is to rely on C++ templating, in practice this approach is often limited and thus supplemented by the use of a domain-specific language such as UFL \cite{ufl} and code generation for hardware-specific optimisations, see e.g. \cite{dune2018, yateto}. 
To solve the contradictory goals of being both an optimized custom-made solver and a broad general-purpose framework, ExaHyPE uses code generation and modularity.
The ExaHyPE engine isolates its most compute intensive routines into modular kernels. 
These kernels are created using code generation to be able to choose the most appropriate numerical schemes for a given application and further tailor them to a given set of requirements.
Code generation also allows the engine to rely on tailored glue code to bind user written functions that implement the desired PDE system, and the suitable kernels to its engine.

The generation of the glue code and kernels is performed by two custom Python~3 modules, the \emph{Toolkit} and the \emph{Kernel Generator}. 
They are designed to be expandable and accommodate for new user requirements.

This interplay of user-provided PDE-specific code, generated glue code and kernels, and hardware-aware optimization of both,
typically requires the combined expertise of interdisciplinary teams. 
We have observed that in such teams the following roles exist and need to be addressed by the engine:

The \textbf{application expert} implements the PDE system for a given application, as well as problem-specific initial and boundary conditions or criteria for mesh refinement and admissibility of solutions.
This role desires a straightforward user API that requires only knowledge about the application and hides the complexity of the solver and its optimizations.
It expects a general-purpose framework with best-possible flexibility in terms of implementing various PDE systems, respective application scenarios and postprocessing of results. 

The \textbf{algorithms expert} tunes the numerical solvers, for performance or numerical properties, 
breaking them down into sequences of kernel calls. 
New algorithmic schemes need to be made available, to expand ExaHyPE's capabilities to
tailor itself toward applications matching specific numerical requirements.
The algorithms expert needs to be able to design these parts in an architecture-oblivious way, while still getting low-level optimizations automatically in the generated code.

The \textbf{optimization expert} contributes architecture specific knowledge and optimizes all performance-critical components of the solver. 
This role requires tools to impact in the background the work done by the other two, while being able to efficiently support multiple architectures.

The three roles might be taken by a single person, but will usually be distributed to teams, such that each role might even be adopted by several persons. While a separation of concerns is used in other PDE frameworks, such as Firedrake \cite{rathgeber2017firedrake, firedrake2} (following a compiler-based approach), we put a special emphasis on the ability of the three experts to extend the code generation itself. 
ExaHyPE thus uses its code generation utilities to isolate the roles, allowing users to focus on their area of expertise, while integrating their cumulative work.
To achieve this, we took inspiration from web application development practices and designed both code generation modules using a Model-View-Controller (MVC) architectural pattern on top of a template engine library, Jinja2 (\url{http://jinja.pocoo.org/}). 
Such template engines are not often used in HPC software -- 
an exception being the MESA-PD particle dynamics code developed within the waLBerla framework \cite{mesapd}. 
There, template logic is used to decouple physical interaction models from the remaining framework. 
For a general PDE framework, such as ExaHyPE, a similar decoupling is not sufficient, however, due to the large number of supported use cases, the use of external libraries and the strong interdependence of the numerical methods. 
%
In this paper, we show how ExaHyPE's code generation utilities and their design choices support the separation of roles and foster optimization of ExaHyPE towards an exascale PDE engine.

We start with a brief overview of the engine, its ADER-DG scheme, kernels and pre-compile-time code generation utilities in the Toolkit and Kernel Generator. 
In Sec.~\ref{sec:codegen} we discuss the architecture of our code generation utilities and how the MVC pattern and Jinja2 are used 
to make code generation and optimization more straighforward.
We then discuss on three use cases how these choices translate into a simplified workflow for each of the three identified roles.

\section{The ExaHyPE engine}

In this section, we provide the numerics of the ADER-DG scheme used by ExaHyPE and how it allows 
to design a framework for solving a wide range of applications.
We then motivate the use of code generation to provide a smooth user experience for the application expert while providing opportunities for algorithms and optimization experts.

\subsection{A high-order ADER-DG solver with a-posteriori limiting}
\label{sec:math_background}
The ExaHyPE engine \cite{Reinarz:2019} can solve a large class of systems of first-order hyperbolic PDEs, which are expressed in the following canonical form:
\begin{equation} \label{eq:hyperbolicPDE}
\frac{\partial \mathbf{Q}}{\partial t}(x,t) + \nabla \cdot \mathbf{F}(\mathbf{Q})
+ \mathbf{B}(\mathbf{Q})\cdot \nabla\mathbf{Q}(x,t) = \mathbf{S}(\mathbf{Q}).
\end{equation}
$\textbf{Q}(x,t) \subset \mathbb{R}^q$ is a space- and time-dependent state vector 
for any $x\in \Omega \subset \mathbb{R}^d$ ($d=2,3$) and $t\in\mathbb{R}_0^+$. 
$\mathbf{F}$ denotes the conserved flux vector, $\mathbf{B}$ the (system) matrix composing the non-conservative fluxes and $\mathbf{S}(\mathbf{Q})$ the source terms. 


To solve equations of this form, ExaHyPE uses the arbitrary high-order accurate ADER Discontinuous Galerkin (DG) method 
in the formulation by Dumbser et al. \cite{Zanotti:2015}. 
The computational domain $\Omega$ 
is discretized with a tree-structured Cartesian grid using the Peano framework \cite{Weinzierl:2019} as mesh infrastructure, allowing for dynamic adaptive mesh refinement.
As understanding the kernels described in the following sections relies on an understanding of the numerical scheme we will briefly sketch the ADER-DG method. More details on the implementation in the engine are given in \cite{charrier2018enclave}.

The ADER-DG method consists of two phases, a predictor step in which the weak formulation of \eqref{eq:hyperbolicPDE} is solved locally in each cell, and a corrector step in which the contributions of neighboring cells are taken into account. 
To derive the weak solution of the problem we insert the DG ansatz function from the space of piecewise polynomials into equation \eqref{eq:hyperbolicPDE} and multiply with a test function from the same space of piecewise polynomials. We then integrate over a space-time control volume. 
The solution of the resulting element-local problem makes up one of our most compute-intensive kernels, the \emph{space-time predictor}.
In non-linear problems the solution of this element-local weak form is calculated using Picard iterations, in the linear setting it can be computed directly using the Cauchy-Kowalewski procedure.

In the second phase the element-local predictor solution is corrected, using contributions from neighboring cells.
 To solve the surface integrals we introduce a classical Riemann solver  as it is used in Godunov-type FV schemes. 
After this correction the next time-step can be calculated. The next time step size depends on the CFL number.

However, this high-order approach suffers from oscillations at shocks and discontinuities. 
We therefore apply an a-posteriori Finite Vollme limiter \cite{Loubere:2013}. 
We identify cells as troubled using the following detection criteria: a relaxed discrete maximum principle in the sense of polynomials, 
absence of floating point errors (NaN, e.g.) and positivity (or similar physical constraints) of the solution. 
If one of these criteria is violated after a time step, the scheme recomputes the solution in the troubled cells, 
using a more robust high resolution shock capturing FV scheme on a subgrid composed of $(2N+1)^d$ cells. 
This procedure is composed of several kernels, the computation of the discrete maximum principle, and a projection from DG to FV solution and vice versa.

ExaHyPE thus provides building blocks to solve specific PDE systems with a tailored scheme: DG vs.\ FV-only, 
DG with or without limiting, Cauchy-Kowalewski procedure or Picard loops for linear or non-linear schemes, 
various choices of Riemann solvers, etc. -- presenting this complexity to users (depending on their roles) 
and thus keeping the engine and derived simulation software manageable are consequently intrinsic challenges for the engine 
development.

\subsection{Application-specific programming interface}

The canonical PDE system \eqref{eq:hyperbolicPDE} can model a wide array of applications,
including relativistic astrophysics \cite{dumbser:axioms,fambri}, seismic wave propagation \cite{SIStag2017,TAVELLI2019} or several variants of fluid equations (see \cite{Reinarz:2019} for an overview).
All these problems can be formulated via Equation \eqref{eq:hyperbolicPDE} via specific $\mathbf{F}(\mathbf{Q})$, $\mathbf{B}(\mathbf{Q})\cdot \nabla\mathbf{Q}$ and $\mathbf{S}(\mathbf{Q})$. 
However, not all these terms occur in every PDE; the engine should therefore not force the user to provide a useless zero function.

Hence, as the first step, the application expert is expected to provide a \emph{specification file} to describe the application and its runtime parameters.
This includes but is not limited to:
\begin{itemize}
\item Application parameters such as the number of quantities in the vector $\mathbf{Q}$ or the polynomial order for the ADER-DG scheme;
\item Which terms of the canonical PDE \eqref{eq:hyperbolicPDE} will be required;
\item Whether the application will require an a-posteriori limiter (as described in Sec.~\ref{sec:math_background});
\item Optimization specific options that can be enabled to further improve the application performance.
\end{itemize}
The specification file relies on a Domain Specific Language (DSL) defined via JSON Schema (\url{https://json-schema.org/}).
Using JSON Schema and its open source tools simplifies both the validation of a given specification file and the modification the DSL (e.g, to introduce new options) as will be described in the use cases.

The specification file is passed to the code generation utilities that set up the engine, and generate glue code and kernels.
This includes a class \lstinline{UserSolver} where the application expert shall implement the required \emph{user functions}:

\begin{itemize}
\item PDE-related functions that provide an implementation of the required terms in \eqref{eq:hyperbolicPDE}, such as a flux function to compute $\textbf{F}(q_h)$;
\item Initial and boundaries conditions;
\item Eigenvalues and physical admissibility;
\item Mesh refinement criteria (if mesh refinement is enabled).
\end{itemize}
The application can then be compiled using a generated Makefile and executed with the specification file as argument for its runtime parameters.

Thus, from an application expert's perspective, 
ExaHyPE allows to solve complicated PDE systems with minimal code writing and without considering the complex issues of designing a performance-oriented high-order solver on a parallel compute cluster.
Hence, ExaHyPE is well suited to quickly build an application for a given PDE system and obtain first insights whether the engine will fit the problem to be tackled.

\subsection{Architecture-aware optimization of kernels}
\label{sec:kernels}

Toolkit and Kernel Generator aim to tailor the engine toward its applications and a target architecture, such that an HPC-worthy production code is produced.
To enable this tailoring, the ExaHyPE engine itself is modular.

This is motivated by the fact that the element-local computation of ADER-DG updates (cf.~Sec.~\ref{sec:math_background}) 
naturally breaks down into substeps (space-time predictor, Riemann solver, etc.), 
which can often again be formulated as smaller substeps (such as tensor or matrix operations).
In the engine, each of these substeps is isolated into a specific kernel. 
These kernels are the critical parts of ExaHyPE, both performance-wise and regarding the implementation of the numerical scheme \cite{charrier2018enclave}.

Having more knowledge about the target application allows more specific but also more efficient numerical schemes, 
such as using a linear scheme instead of a general nonlinear one. 
Likewise, a specific numerical scheme may be required to satisfy certain stability constraints, such as using a special Riemann solver.
Finally, knowing the target architecture enables different low-level optimization techniques, such as the supported SIMD features and the required array alignment and padding settings.

The Kernel Generator uses all information provided in the specification file to choose the correct scheme for each kernel and uses code generation to add application- and architecture-aware optimizations to them.
Using code generation also facilitates the inclusion of external performance related libraries and code generators.
The generated kernels are bound to the engine core using the Toolkit's generated glue code.

We cannot expect that the set of alternative schemes and supported optimizations provided by the Kernel Generator will ever be complete. 
New user and hardware requirements will arise constantly. 
Therefore, to facilitate the work of algorithm and optimization experts, 
the Kernel Generator is designed with ease of modification in mind, so that they can enrich the available customization options of the engine.
As adding new options for the Kernel Generator translates into expanding the specification file DSL and adapting the glue code, the Toolkit's design follows the same philosophy.

\section{Code Generation in ExaHyPE} \label{sec:codegen}

%

\subsection{Model-View-Controller Design}
\label{sec:codegen_struct}

The Toolkit and the Kernel Generator are implemented as Python~3 modules.
Python was chosen for its ease of use and development, as well as for its mature open source ecosystem.
%
Both modules follow the Model-View-Controller (MVC) architectural pattern, which is widely used, especially in web applications.
Our motivation toward using an MVC pattern is twofold. 
First, the goal of generating user-tailored HTML pages and building an application by combining multiple separate developer roles is quite similar to our own situation.
MVC has managed to become an industry standard, being recognized for the ease of development, code reusability and useful abstraction layers it provides.
Second, we can re-purpose mature open source tools, such as the Jinja2 template engine, to generate C++ code instead of its intended HTML output.
Using a template engine allows us to streamline the development of new features and to separate the implementation of a new numerical scheme to its low-level optimization.

Reformulated in the MVC paradigm, each of our desired C++ files to be generated (kernel and glue code alike) is a \emph{View} to be rendered by a \emph{Model} responsible for it 
and the specification file is the input of the \emph{Controller}.
The Toolkit implements the MVC pattern in the following way:

\paragraph{\bfseries Controller} The Toolkit's Controller class validates the specification file, parses it, and builds multiple contexts, implemented by Python dictionaries.
Each context contains only the relevant information for a given Model, thus providing an abstraction layer between the specification file grammar and the internal Toolkit API.
The Controller calls the application relevant models only and passes them their respective context.
For example a Python dictionary containing the application name, path and target architecture (if provided) is generated and passed as context to the Model responsible for generating the \lstinline{Makefile},
while the Model responsible for building the \lstinline{UserSolver} contains the solver relevant information, such as the polynomial order of the ADER-DG scheme or the used terms of the canonical PDE form \eqref{eq:hyperbolicPDE}.


\paragraph{\bfseries Model} Each Model is responsible for generating a specific View, or group of Views.
After receiving its context from the Controller, a Model may expand it using its own internal logic to add relevant internal parameters.
In situations where different versions of a View exist, it decides which one is required.
For example, it might choose a View to generate the glue code for either a finite volume solver or an ADER-DG solver, which require different kernels.
It selects the appropriate template that represents the desired View version, or in a simpler case uses the sole template for this View.


\paragraph{\bfseries View} Views are implemented by templates which are a generalized representation of a given C++ code that may be tailored to a specific context. 
The Jinja2 template engine is invoked to render a template with a Model-provided context.
Jinja2 parses its input template and uses the context to interpret it.
Its output is then written as a valid C++ file that matches the context, and thus specification file, requirements. 
For example, it may hard-code the selected polynomial order and use the generated kernels.

\bigskip

The Kernel Generator follows the same MVC architecture and is called by a special Model of the Toolkit.
This Model translates its context into the required format for the Kernel Generator API and passes it to its Controller. The same MVC schema is then replicated.
The separation of Toolkit and Kernel Generator into two utilities is dictated by their different purpose: 
The Toolkit generates glue code and code the application expert is expected to interact with, 
while the Kernel Generator handles numerical schemes and low-level optimizations for the other two roles.


\subsection{Templates}
\label{sec:template}

As mentioned in Sec.~\ref{sec:codegen_struct}, a template is a generalized representation of a given C++ file that we want the code generation utilities to generate -- e.g., a kernel or some glue code.
By using templates, we are able to put some logic in the code representation while keeping it close to the generated code and thus easily readable and expendable.

To express this logic, we use the templating language implemented by Jinja2.
Its language syntax is designed to be both easy to learn and to work with, and is therefore well suited to allow ExaHyPE's users to modify the behavior of its code generation utilities.
It also provides some advanced functionalities that can be used directly in the code abstraction.


\begin{figure}[tbp]
\begin{lstlisting}[basicstyle=\ttfamily\small]
// template
{% if initA %}
{{allocateArray('A', nDof)}}
for(int i=0; i<{{nDof}}; ++i) {
  A[i] = B[i+{{nDof*nVar}}] * {{C}}[i];
}
{% endif %}

// generated code
double A[5] __attribute__((aligned(32)));
for(int i=0; i<5; ++i) {
  A[i] = B[i+20] * foo[i];
}
\end{lstlisting}
\caption{Example of a template and the resulting generated code\label{fig:template_frag}}
\end{figure}

The code fragment in Fig.~\ref{fig:template_frag} illustrates how we use templates to generate C++ code:
At its simplest any given string or number can be abstracted behind a variable in a template's context. 
This is used, for example, to abstract the application's namespace, which depends on the user specification, in the template.
Mathematical computations can also be done and the result directly written in the generated code, 
In the code of Fig.~\ref{fig:template_frag} this is used to hard-code the loop boundary \lstinline{nDof}, the index shift of the array \lstinline{B} and the name of the third array.

Furthermore, boolean operations and branchings are used to selectively enable or disable certain parts of the generated code.
For example in the glue code responsible for binding the kernels to the engine, choosing linear or nonlinear kernels is done using Jinja2's branching.
This allows us to efficiently deal with the multitude of options ExaHyPE offers its users, without having to duplicate code or use slower runtime branching.
In the code of Fig.~\ref{fig:template_frag}, branching is used to include the whole fragment only if the context's boolean \lstinline{initA} is true.

Jinja2's logic also includes subtemplating, i.e. including and rendering a template inside another one, and custom macros.
With this we can factorize repeating portions of the templates, thus making them easier to maintain and expand.
We also use macros to provide architecture-aware optimizations.
In Fig.~\ref{fig:template_frag}, we use the macro \lstinline{allocateArray} to allocate a new array \lstinline{A}.
This macro abstracts the optimized allocation of an array of a given size.
In our example, it produces the C++ code to allocate the array \lstinline{A} on the stack and on a 32-bytes boundary for more efficient AVX2 operations. 

\subsection{Architecture-oblivious templates and architecture-aware optimization macros}
\label{sec:template_opt}

\begin{figure*}[h]
\begin{lstlisting}[basicstyle=\ttfamily\small]
{% macro allocateArray(name, size, setToZero=False) %}
{% if tempVarsOnStack %}
double {{name}}[{{size}}] __attribute__((aligned(       \
    {{alignment}}))) {{"={0.}" if setToZero}} ;
{% else %}
double* {{name}} = ((double*) _mm_malloc(sizeof(double) \
    *{{size}}, {{alignment}}));
{% if setToZero %}
std::memset({{name}}, 0, sizeof(double)*{{size}});
{% endif %}
{% endif %}
{% endmacro %}
\end{lstlisting}
\caption{Example of an optimization macro to allocate arrays\label{fig:alloc_macro}}
\end{figure*}

The Kernel Generator provides kernels that are optimized toward both given application requirements and a target architecture.
The former is done via algorithmic adaptations: choosing the appropriate scheme, enabling or disabling features, hard-coding specific values, etc.
The latter requires low-level code optimizations (e.g., array padding and alignment), compiler specific pragmas and instructions, or external libraries.
Performing both at the same time on a given kernel template would make it hard to read, maintain and expand. 
Hence the separation of the role of algorithms and optimization experts.

Using Jinja2's macros and variables, we can design an architecture-oblivious template that will be rendered with architecture-aware optimizations.
Thus, most templates in the Kernel Generator are \emph{algorithmic templates}: templates that focus on describing a given scheme with some algorithmic optimizations but without any complex logic for architecture related ones.
A second smaller set of templates define \emph{optimization macros} and the subtemplates used by these macros to perform a specific task or output a specific architecture-aware optimization.
The macros defined this way can then be used by the algorithmic templates. 

The code excerpt in Fig.~\ref{fig:alloc_macro} shows a simplified version of the \lstinline{allocateArray} macro that was used in Fig.~\ref{fig:template_frag}. 
It takes the array's \lstinline{name} and \lstinline{size} as positional inputs and optionally a boolean \lstinline{setToZero} to indicate if the array should be initialized to zero.
Then, depending on a global optimization flag \lstinline{tempVarsOnStack}, it allocates the array either on the stack or on the heap.
Enabling this feature depends on the target hardware setting, as a limited stack size could cause crashes.
The \lstinline{allocateArray} macro takes care of array alignment to optimize for SIMD using a global \lstinline{alignment} context parameter 
that is set by the Kernel Generator's Controller depending on the specified target architecture, and thus the target AVX settings.
For heap allocation, a compiler-specific instruction is used (e.g., \lstinline{_mm_malloc} for the Intel compiler).

Thus every time a temporary array is needed, it can be allocated using this macro, hiding the low-level optimization from the algorithms expert.
If the optimization expert needs to add support for a different compiler, e.g., expanding this macro provides it to all kernels.
A complementing \lstinline{freeArray} macro exists to free the memory correctly, as for example using \lstinline{_mm_malloc} requires using the Intel-specific \lstinline{_mm_free} instruction, 
whereas the pointer should not be freed at all, if a stack allocation was used.

Macros can also be used to include external libraries.
For example ExaHyPE's kernels spend a lot of computational effort in performing small dense matrix products that result from expanding respective element-local tensor operations.
For these we employ LIBXSMM \cite{libxsmm}, which generates architecture specific function kernels to perform small matrix products at best-possible performance on a given Intel architecture.
Using a custom \lstinline{matmul} macro and with some modification to the controller and models to properly define the parameters of each matrix products in the template, 
LIBXSMM can be selected and integrated into the kernels.
By expanding the \lstinline{matmul} macro, an optimization expert can also easily switch to another library to support another kind of architecture.

Thus the development of new numerical schemes and the low-level architecture-aware optimization can be kept separated.
This ensures that the role of algorithm and optimization expert are independent from one another.

\section{Expanding the PDE: Navier-Stokes equations}
\label{sec:mod_kernel}
In this section, we discuss the solution of the compressible Navier-Stokes equations using the ExaHyPE engine~\cite{Reinarz:2019}. 
Following our PDE system \eqref{eq:hyperbolicPDE}, we can write the compressible Navier-Stokes equations as

\newcommand{\Q}{\mathbf{Q}}
\newcommand{\gradQ}{\nabla\Q}
{\small
\begin{equation}\label{eq:navier-stokes}
\frac{\partial}{\partial t} \!
\underbrace{
\begin{pmatrix}
  \rho\\
  \rho v \\
  \rho E
\end{pmatrix}}_{=\Q} + \nabla \cdot
\underbrace{\begin{pmatrix}
\rho v\\
v \otimes \rho v + I p + \sigma (\Q, \gradQ)\\
v \cdot \left( I \rho E + I p + \sigma(\Q, \gradQ) \right)
 - \kappa \nabla (T)
\end{pmatrix}}_{=\mathbf{F}(\Q, \gradQ)}
=
S(Q).
\end{equation}
}\noindent
where $\rho$ denotes the density, $\rho v$ the momentum, $\rho E$ the energy density, $T$ the temperature and $p$ the pressure (including hydrostatic pressure, e.g., gravitational effects). 
The temperature diffusion is given by $\kappa \nabla T$ with constant $\kappa$.
Note that the stress tensor $\sigma(\Q, \gradQ)$ involves a parabolic component, expressed via the dependence on $\gradQ$. 
While we can largely stay with the existing numerical approach to solve the equations in ExaHyPE, we had to extend the API to allow for flux terms $\mathbf{F}(\Q, \gradQ)$ that depend on $\gradQ$ in the canonical PDE \eqref{eq:hyperbolicPDE}. 
For example, the gradient of the state vector $\Q$ had to be added as argument of the flux function, which was also renamed to \lstinline{viscousFlux}.
Further changes are modifications to the space-time predictor, the boundary conditions and the introduction of a new Riemann solver~\cite{Gassner:08:ViscousFlux}. 

As only minor modifications to the existing numerical schemes and none to the optimizations are required, we followed a straightforward linear workflow:

\subsection{Expanding the DSL}
We modified the DSL of our specification file to include a new optional flag enabling the viscous flux terms in the PDE system as an opt-in feature.
In the JSON schema, this meant adding a \lstinline{viscous_flux} option to the already existing list of optional PDE components for an ADER-DG Solver:


\begin{lstlisting}[basicstyle=\ttfamily\small]
"items":{ 
 "type":"string", 
 "enum":["flux","source","ncp","viscous_flux"]
}
\end{lstlisting}

The Schema processing library only performs basic input validation. 
Here \lstinline{flux} and \lstinline{viscous_flux} should not appear together, 
thus a new test was added to the validation method of the Toolkit's Controller class, such that an error message is issued if a user selects both options simultaneously.

\subsection{Processing the new specification file option}
The new \lstinline{viscous_flux} option is processed by the Controller and passed on as a boolean flag \lstinline{useViscousFlux} in the context of the Models needing to act on it.
In the MVC architecture, the addition of a few lines of code is sufficient to provide the Views with such a boolean flag.

\subsection{Expanding the Views}

The code to be generated is abstracted in the Views of the Toolkit and Kernel Generator by Jinja2's templates.
Using Jinja2'2 template branching logic, 
the application expert is asked to provide a \lstinline{viscousFlux} user function in the generated UserSolver,
if the \lstinline{useViscousFlux} flag is set. 
Then in all kernels using \lstinline{flux}, the gradient $\gradQ$ is computed using already existing macros, which deal with the optimization of this computation, 
and the \lstinline{viscousFlux} function is called instead with it as additional argument.

The branching also ensured that the expanded Views generate the same code as before if the flag is not set (opt-in option). 
Since every part of the code generation is compartmentalized into separate Models, modifying a Model or expanding the Controller has no side effects on the other generated code.

\subsection{Result evaluation}
%
%
%
%
%
%

An application using this new feature was written and tested.
It is able to simulate cloud formation processes in scenarios incorporating a background atmosphere that is in hydrostatic balance~\cite{Krenz:19:ExaCloud}.  
At the end of this use case, the ExaHyPE engine's canonical PDE system \eqref{eq:hyperbolicPDE} is expanded and can now, as an opt-in option, 
work with further applications requiring a viscous flux term instead of a classical flux. 

The modifications needed to implement the features required roughly 100 lines for the kernels and additionally less than 100 lines for the Toolkit.
This includes all code, comments and all needed API changes.
We want to emphasize that theses changes required only a basic algorithmic understanding and minimal optimization knowledge, thanks to the reuse of existing optimization macros.




\section{Improved space-time predictor for linear applications}



Benchmarks of the linear PDE solver at high polynomial orders revealed significant loss of performances due to cache misses inside the SpaceTimePredictor kernel.
This was caused by the temporary arrays required by the algorithm to implement the Cauchy Kowalewski scheme inside this kernel.
The size of these arrays depends on the polynomial order used, and increased beyond the L2 cache size of our test hardware during benchmarks.
Thus, to reduce the memory footprint, we reformulated the algorithm toward cache efficiency.

Instead of storing all time derivatives for later integration, the time integration is performed on the fly. 
Thus the full time dimension is removed from temporary arrays.
As a result, the spatial directions of the PDE system are processed one at a time.
The algorithm therefore requires three directional \lstinline{flux} functions instead of one for all dimensions.
Depending on the application specific formulation this might lead to redundant computations.
Therefore, despite being more memory efficient, the new algorithm is offered as an optional kernel variant (opt-in option).
To introduce this new SpaceTimePredictor kernel variant, we used an iterative and incremental approach:

\subsection{Prototyping the new algorithm} 
The new algorithm was first prototyped on a test application with fixed settings.
We generated the default SpaceTimePredictor kernel for the test case and edited it locally to get to the new algorithm.
This way we could test the new algorithm, verify it against the default one and validate our assumption on improving the memory footprint.
We then iterated upon the prototype to incrementally add new optimizations, as tests revealed bottlenecks and possible areas of improvements.

\subsection{Inclusion in the Kernel Generator} 
Once the prototype was finished and validated, 
it was incorporated directly into the Kernel Generator.
The prototype source code was directly used as the first iteration of a new template,
since a template can also exist of explicit code without any template logic.
Then using the existing MVC structure of the Kernel Generator, its generation behavior was modified by introducing a new optional input parameter to trigger the generation of this new template 
(as in the use case of Sec.~\ref{sec:mod_kernel}).
At this stage the Kernel Generator was able to generate the prototype kernel variant only for the application and setting it was designed for during the prototyping step.

\subsection{Template generalization and optimization} 
Finally the template was generalized, such that it can be used with other settings or by other applications.
The hard-coded settings from the prototyping steps (e.g., the name of the solver, the polynomial order) 
were replaced by their respective abstractions, as defined in the provided template context, thus enabling the new kernel variant to be properly generated for all settings.
This transformed the prototype template to an algorithmic template as described in Sec.~\ref{sec:template_opt}.

To provide architecture-aware optimizations, we used the existing optimization macros, for example to perform optimized matrix products.
Thus, this new kernel variant was immediately optimized toward all the supported architecture without needing any optimization knowledge.

\subsection{Performance evaluation}

Once the new kernel variant was fully supported by the code generation utilities, we used ExaHyPE's internal benchmarking tools to compare it with the default one 
on a set of test applications, settings and architectures.
These tests confirmed our early intuition that the new algorithm provides no runtime benefits for applications with low memory footprint, 
but leads to speedups of $>2$ for bigger settings that are severely affected by cache misses with the default algorithm.
The threshold depends on the application, its settings (esp.\ the polynomial order) and the hardware specification (esp.\ the L2 cache size).

Using the Kernel Generator MVC architecture and the optimization macro, the development of this new kernel variant, from building a prototype to the benchmarking of the feature,
required almost exclusively the numeric and algorithmic optimization expertise, expected from an algorithm expert role.


\section{Vectorization of user functions}
\label{sec:userfun_vect}

The last use case addresses the exploitation of SIMD capabilities of modern CPUs. 
Here, ExaHyPE faces a conflict of API and optimization requirements. 
For the implementation of user functions, such as the flux function $\mathbf{F}(\mathbf{Q})$, 
the most intuitive API is like the function \lstinline{flux(Q,F)} in Fig.~\ref{fig:vectPDE}: 
\lstinline{flux} acts on a contiguous vector of quantities.
This Array of Structure (AoS) data layout also supports the optimized execution of 4D tensor operations 
(3D space plus the quantity dimension) via sequences of matrix operations -- 
the matrices always have the quantities as a dimension that is contiguous in memory. 
However, AoS becomes inefficient, when calling the user functions for multiple spatial positions, 
such as evaluating the flux function at all integration points to evaluate the Riemann problem on element faces. 
The kernels then loop over all spatial coordinates, but call the user functions on the vector $\Q$ for each single spatial point.
These calls cannot be vectorized, as the accessed components \lstinline{Q[0]}, \lstinline{Q[1]}, etc. 
(similar for \lstinline{F[0][0]}, \dots) are not stored in unit-stride.

To solve this data layout conflict, we introduced SIMD user functions as opt-in features.
Instead of processing a single quantity vector, they take as parameter a vector of quantities in a Structure of Array (SoA) layout, so that the resulting loop in the implementation can be vectorized.
Fig.~\ref{fig:vectPDE} illustrates how to implement such a SIMD flux function (\lstinline{fluxVect}): 
The input arrays of \lstinline{fluxVect} now have a new fastest-running dimension that matches the loop iteration, 
such that compiler auto-vectorization may be enabled. 

\begin{figure}[h]
\begin{lstlisting}[basicstyle=\ttfamily\small]
void Euler::flux(double* Q, double** F) {
  //[...] constants
  // x direction
  F[0][0] = Q[1];
  F[0][1] = irho*Q[1]*Q[1] + p;
  F[0][2] = irho*Q[2]*Q[1];
  F[0][3] = irho*(Q[3]+p)*Q[1];
  //[...] y direction
}

void Euler::fluxVect(double** Q, double*** F){
  #pragma vector aligned
  #pragma ivdep
  for(int i=0; i<VECTSIZE; i++){
    //[...] constants
    // x direction
    F[0][0][i] = Q[1][i];
    F[0][1][i] = irho*Q[1][i]*Q[1][i] + p;
    F[0][2][i] = irho*Q[2][i]*Q[1][i];
    F[0][3][i] = irho*(Q[3][i]+p)*Q[1][i];
    //[...] y direction
  }
}
\end{lstlisting}
\caption{Example implementation (for the 2D Euler equations) of a flux function $\mathbf{F}(\mathbf{Q})$ for a single flux vector $\Q$ 
   (\lstinline{flux(...)}, top) or for an array of flux vectors (\lstinline{fluxVect(...)}, bottom).
   Note that in \lstinline{F[0][0][i]}, etc., \lstinline{i} is the fastest-running index.
   \label{fig:vectPDE}}
\end{figure}


In this use case, we describe the integration of these new user functions to all existing kernels using new optimization macros.
By using macros, only optimization specific knowledge is required during development and they can be reused by algorithms experts when implementing new schemes.
We will describe only the work for the \lstinline{flux} user function, the same being done for the others.

\subsection{Optimized transpose -- from AoS to SoA (and back)}

To be able to use a SIMD user function, the data layout has to be transformed on the fly from AoS to SoA and back.
This is achieved by transposing a slice of the input array to a new temporary array.
Processing with slices instead of the whole array optimizes caching behaviors.

We therefore introduced a new optimization macro called \lstinline{transpose}.
By default it falls back to a naive loop-based transpose.
However, more optimized transpose implementations are offered, such as ones using architecture-specific intrinsic operations like
 \lstinline{_mm256_permute2f128_pd} and \lstinline{_mm256_shuffle_pd} for AVX.
At rendering, the best available implementation for the given context is chosen.
It can easily be expanded to better support other architectures and could be expanded to use an external library like the matrix product \lstinline{matmul} macro with LIBXSMM.

\subsection{Abstracting the call to the user function behind a macro}

The choice between the \lstinline{flux} and \lstinline{fluxVect} user functions and the required supporting logic is complex and repeated at each instance where the flux function $\mathbf{F}(\mathbf{Q})$ is evaluated in the kernels.
As described in Sec.~\ref{sec:template_opt}, we can factorize this template code and abstract it behind a new optimization macro named \lstinline{callFlux}.
We started by abstracting the current behavior behind the \lstinline{callFlux} macro:

\begin{lstlisting}[basicstyle=\ttfamily\small]
{% macro callFlux(Q, F, size) %}
{% set F_shift = nDof**nDim*size %}
double* F[{{nDim}}];
for (int i = 0; i < {{nDof**nDim}}; i++) {
  F[0] = {{F}}+i*{{size}};
  F[1] = {{F}}+i*{{size}}+{{F_shift}};
{% if nDim == 3 %}
  F[2] = {{F}}+i*{{size}}+2*{{F_shift}};
{% endif %}
  {{solverName}}.flux({{Q}}+i*{{size}},F);
}
{% endmacro %}
\end{lstlisting}

At that point it performed only the existing default case to call the \lstinline{flux} function: 
loop over all spatial points of the cell, initialize the array \lstinline{F} and call the function with the correct shift in the data arrays as they use an AoS layout.
The evaluations of the flux function in all kernels are replaced by \lstinline{callFlux}.

\subsection{Expanding the \lstinline{callFlux} macro}

As in the two previous use cases, we introduced a new context boolean flag \lstinline{useFluxVect}.
We then expanded the \lstinline{callFlux} macro with a branch on this flag. 
If the flag is set, the \lstinline{callFlux} macro uses the \lstinline{transpose} macro defined earlier to switch on the fly between AoS and SoA data layout 
and call a new \lstinline{fluxVect} user function with its altered signature compared to \lstinline{flux} as shown in Fig.~\ref{fig:vectPDE}.
As we modified a macro, this work is automatically propagated to all existing templates using it.

\subsection{Performance evaluation}

We evaluated the SIMD user functions on two example PDEs: the 3D Euler equations (EulerFlow), where the flux function is quite simple, and the Einstein equations from relativistic astrophysics (CCZ4), where the user functions are highly complex and comprise most of the runtime.
The benchmark was done on SuperMUC phase 2 (Intel Haswell architecture, supporting AVX2).
For EulerFlow we compared the default version with the auto-vectorized one and with an intrinsics-version for AVX2.
The auto-vectorized and the intrinsics implementations both achieved similar performances, illustrating that for simple user functions a quick adaptation of the scalar implementation to enable auto-vectorization is enough.
Compared to the default version, both SIMD implementations provided an end-to-end speedup by a factor $1.04$.
Here the low cost of the simple user function is barely enough to compensate for the cost of the required transpositions.

With the help of an application expert, we implemented a partially auto-vectorized version of the complex CCZ4 user functions. We measured a speedup factor of $1.27$.
While the user functions were not fully vectorized due to their complexity, their high computational cost is enough to offset the transpose one.
A better vectorized implementation of the user functions would provide even more performance gain.

Here by working with macros, we not only provide these new features to all existing schemes, but also ensure that future ones can easily use them.
The implementation of the macros required mostly low-level optimization knowledge. 
All architecture-specific optimizations are fully handled by the macros, enabling an optimization expert to easily improve them or expand them for other architectures.


\section{Conclusions}

This paper details how code generation is used in a PDE engine 
to offer a tailored application-specific programming interface for users, 
while at the same time selecting the most appropriate (regarding the target application) 
numerical scheme and implementation for each of its critical components,  
and tuning it with low-level architecture-aware optimizations.
The choice of a MVC architecture for code generation facilitates the collaboration of three 
identified user roles -- application, algorithm and optimization experts -- as they use and expand the engine.
In the Views, Jinja2's template logic and macros support
the implementation of new algorithms and low-level code optimization independently of each other.

The three presented use cases show how a user assuming only one single role can work with the engine and contribute to it by expanding the code generation utilities, cumulatively improving its capabilities.
Thus, the presented design solves a common issue encountered when building complex HPC simulation software: to support users with different areas of expertise in their effective collaboration.



\section{Acknowledgements and Funding}

This project has received funding from the European Union's Horizon 2020 research and innovation programme under grant agreement No 671698.
We thank the Gauss Centre for Supercomputing e.V. (\url{www.gauss-centre.eu}) for providing computing resources on the GCS Supercomputer SuperMUC at Leibniz Supercomputing Centre (\url{www.lrz.de}).


\bibliographystyle{splncs04}
\bibliography{references}



\end{document}